\documentstyle[12pt]{article}

\def\be{\begin{equation}}
\def\ee{\end{equation}}
\def\ba{\begin{array}{c}}
\def\ea{\end{array}}

\def\ben{$$}
\def\een{$$}
\begin{document}

\titlepage

\begin{center}{\Large \bf
  ${\cal PT}-$symmetric form of the Hulth\'en potential
}\end{center}

\vspace{5mm}

\begin{center}

{\bf Miloslav Znojil} \vspace{3mm}

Theory Group,
Nuclear Physics Institute AS CR,
\\CS 250 68 \v{R}e\v{z},
Czech Republic\\

\vspace{5mm}

\end{center}

\section*{Abstract}

The fourth, missing example of an exactly solvable complex
potential with ${\cal PT}$ symmetry $V(x) = [V(-x)]^*$ defined on
a bent contour and leading, at the real energies, to the Jacobi
polynomial wave functions is found in a generalized Hulth\'{e}n
interaction.

\vspace{9mm}

\noindent
 PACS 03.65.Ge,
03.65.Fd


\newpage

\noindent Quantum mechanics is often forced to formulate its
predictions numerically. Its exactly solvable models are scarce.
Only their subclass in one dimension is broader and, in this
sense, privileged and exceptional. It involves the harmonic
oscillator and Morse potentials (with bound states expressible in
terms of the Laguerre polynomials) as well as several models
solvable in terms of the polynomials of Jacobi (cf., e.g., review
\cite{Cooper} for more details).

Within the framework of an alternative or modified quantum
mechanics as proposed recently by Bessis \cite{Bessis} and by
Bender et al \cite{Benderori,Bender} one works with the ${\cal
PT}$ symmetric complex potentials $V(x) = [V(-x)]^*$. The real
bound state spectrum and even the solvability of several real
potentials proves rather unexpectedly preserved after a
complexification of this type. For example, within the
Laguerre-related exactly solvable sub-family we may find both the
complex harmonic oscillator \cite{Benderori} and a complexified
Morse interaction \cite{Morse}. The same high degree of analogy
between the real and complex forces is also observed for the
regular solvable models based on the use of the Jacobi polynomials
\cite{Bagchi}.

A word of warning against unlimited optimism comes from singular
interactions. In particular, for the Laguerre-related and
phenomenologically most appealing Coulombic interaction the only
available ${\cal PT}$ symmetrization proves merely partially
solvable \cite{Coulomb}. In the present communication we intend to
offer a partial remedy. We shall derive and describe a complete
and exact solution for an appropriate ${\cal PT}$ symmetric
complexification of the singular phenomenological Hulth\'{e}n
potential which is known to mimick very well the shape of the
Coulombic force in the vicinity of its singularity \cite{Hulthen}.


In the first step let us recollect that in the spirit of the old
Liouville's paper \cite{Liouville} the change of the (real)
coordinates (say, $r \longleftrightarrow \xi$) in Schr\"{o}dinger
equation
 \be
\left[-\,\frac{d^2}{dr^2} + W(r)\right]\, \chi(r) = -\kappa^2
\,\chi(r)
 \label{SEor}
  \ee
may sometimes mediate a transition between two different
potentials.  It is easy to show \cite{Olver} that once we forget
about boundary conditions one simply has to demand the existence
of the invertible function $r=r(\xi)$ and its few derivatives
$r'(\xi), \, r''(\xi), \ldots$ in order to get the explicit
correspondence between the two bound state problems, viz.,
original eq. (\ref{SEor}) and the new Schr\"{o}dinger equation
  \be
\left[-\,\frac{d^2}{d\xi^2} + V(\xi)\right]\, \Psi(\xi) = E
\,\Psi(\xi)
 \label{SE}
  \ee
with the wave functions
  \be
\Psi(\xi)={ \chi[r(\xi)] / \sqrt{r'(\xi)}} \label{trenky}
  \ee
generated by a new interaction at new energies,
 \be
 V(\xi)-E=\left [
 r'(\xi)
 \right ]^2
 \left \{
 W[r(\xi)]+\kappa^2
 \right \}
+
 \frac{3}{4}
 \left [
 {r''(\xi) \over r'(\xi)}
 \right ]^2
 -
 \frac{1}{2}
 \left [
 {r'''(\xi) \over r'(\xi)}
 \right ].
 \label{newpot}
  \ee
The mapping between the Morse and harmonic oscillators is one of
the best known explicit illustrations of this rule since the
necessary preservation of the correct physical boundary conditions
is very straightforward to check there \cite{Newton}. An
appropriate ${\cal PT}$ symmetrized extension of this equivalence
to the complex non-Hermitian cases is easy \cite{Morse}.

In the alternative, Jacobi-polynomial context the Liouvillean
changes of variables have been applied systematically to all the
Hermitian models (cf. Figure 5.1 in the review \cite{Cooper} or
refs. \cite{Varshni} for a more detailed illustration).  A similar
thorough study is still missing for the ${\cal PT}$ symmetric
models within the same subclass.

In the present letter we shall try to fill the gap.  For the sake
of brevity we shall only restrict our attention to the ${\cal PT}$
symmetric initial eq. (\ref{SEor}) with the P\"{o}schl-Teller
potential studied and solved exactly in our recent preprint
\cite{PTP},
  \be
 W(r)=\frac{ \beta^2-1/4}{\sinh^2 r}
 -\frac{\alpha^2-1/4}{\cosh^2 r}, \ \ \ \ \ \ r = x -
 i\varepsilon,
 \ \ \ \ \ \ \ \ \ x \in (-\infty, \infty)
\label{PTex}
   \ee
The normalizable ${\cal PT}$ symmetric solutions
  \ben
\chi(r) = \sinh^{\tau  \beta+1/2}r
 \cosh^{\sigma\alpha+1/2}r\,\
  P_n^{(\tau  \beta,\sigma\alpha)}(\cosh 2r)
   \een
are proportional to the Jacobi polynomials at all the negative
energies $-\kappa^2 < 0$ such that
  \ben
 \kappa=
 \kappa^{(\sigma,\tau)}_n=-\sigma\alpha-\tau \beta -2n-1>0.
  \een
These bound states are numbered by $n = 0,1,\ldots, n_{max}^{
(\sigma,\tau)}$ and by the generalized parities $\sigma=\pm 1$
and $ \tau =\pm 1$.

We may note that our initial ${\cal PT}$ symmetric model
(\ref{SEor}) remains manifestly regular provided only that its
constant downward shift of the coordinates $r = r_{(x)} = x -
i\,\varepsilon$ remains constrained to a finite interval,
$\varepsilon \in (0,\pi/2)$.  In this sense our initial model
(\ref{PTex}) is closely similar to the shifted harmonic
oscillator.  At the same time, one still misses an analogue of a
transition to its Morse-like final partner $V(\xi)$ in eq
(\ref{SE}).  In a key step of its present construction let us
first pick up the following specific change of the axis of
coordinates,
  \be
 \sinh r_{(x)} (\xi)= - i e^{i\xi}, \ \ \ \ \ \ \ \ \xi = v - iu.
\label{tren}
  \ee
The main motivation of such a tentative assignment lies in the
related shift and removal of the singularity (sitting at $r=0$)
to infinity ($u \to +\infty$). In fact, one cannot proceed
sufficiently easily in an opposite direction, i.e., from a
choice of a realistic $V(\xi)$ to a re-constructed $r(\xi)$.
This is due to the definition (\ref{newpot}) containing the
third derivatives and, hence, too complicated to solve.

We shall see below that we are quite lucky with our purely trial
and error choice of eq. (\ref{tren}). Firstly, we already clearly
see that the real line of $x$ becomes mapped upon a manifestly
${\cal PT}$ symmetric curve $\xi = v - iu$ in accordance with the
compact and invertible trigonometric rules
  \ben
 \sinh x \cos \varepsilon = e^u\,\sin v,\ \ \ \ \ \ \ \ \ \ \ \ \
 \cosh x \sin \varepsilon = e^u\,\cos v,
  \een
i.e., in such a way that
  \ben
 \ba
 v =\arctan \left (
\frac{\tanh x}{\tan \varepsilon} \right )= v_{(x)} \in \left
(v_{(-\infty)}, v_{(\infty)}\right )\equiv \left
(-\frac{\pi}{2}+\varepsilon, \frac{\pi}{2}-\varepsilon \right ),\\
u = u_{(x)} = \frac{1}{2} \ln \left ( \sinh^2x+\sin^2\varepsilon
 \right ) .
 \ea
  \een
This relationship is not too different from its Morse-harmonic
predecessor studied in ref. \cite{Morse}. Our present path of $\xi
$ is a very similar down-bent arch which starts in its left
imaginary minus infinity, ends in its right imaginary minus
infinity while its top lies at $x=v=0$ and $-u=-u_{(0)}= \ln
1/\sin \varepsilon>0$. The top may move towards the singularity in
a way mimicked by the diminishing shift $\varepsilon \to 0$.
Indeed, although the singularity originally occurred at the finite
value $r\to 0$, it has now been removed upwards, i.e., in the
direction of $-u \to +\infty$.


The first consequence of our particular change of variables
(\ref{tren}) is that it does not change the asymptotics of the
wave functions. As long as $r'(\xi)= i\tanh r(\xi)$ the
transition from eq. (\ref{SEor}) to (\ref{SE}) introduces just
an inessential phase factor in $\Psi(\xi)$. This implies that
the normalizability (at a physical energy) as well as its
violations (off the discrete spectrum) are both in a one-to-one
correspondence.

The explicit relation between the old and new energies and
couplings is not too complicated. Patient computations reveal
its closed form. With a bit of luck, the solution proves
non-numerical.  The new form of the potential and of its binding
energies is derived by the mere insertion in eq. (\ref{newpot}),
  \be
 V(\xi)=
 \frac{A}{(1-e^{2i\xi})^2}+\frac{B}{1-e^{2i\xi}},
\ \ \ \ \ \  E=\kappa^2.
 \label{hulth}
  \ee
At the imaginary $\xi$ and vanishing $A=0$ this interaction
coincides with the Hulth\'en potential.

In the new formula one has to notice the positivity of the
energies. It is extremely interesting since the potential itself
is asymptotically vanishing. One may immediately recollect that a
similar paradox has already been observed in a few other ${\cal
PT}$ symmetric models \cite{Morse,quartic,Sesma} where even an
asymptotic decrease of the potential to minus infinity did not
destroy a lower bound of the spectrum.

The exact solvability of our modified Hulth\'en potential is not
yet guaranteed at all. A critical point is that the new couplings
depend on the old energies and, hence, on the discrete quantum
numbers $n$, $\sigma$ and $\tau$ in principle.  This could induce
an undesirable state-dependence into our new potential. Vice
versa, the closed solvability of the constraint which forbids this
state-dependence will be equivalent to the solvability at last.
Fortunately, a smooth removal of the obstacle is possible by a
transfer of the state-dependence (i.e., of the $n-$, $\sigma-$ and
$\tau-$dependence) in
 \ben A=A(\alpha) = 1 - \alpha^2, \ \ \ \ \
\ \ \ C \ (= A +B) =\kappa^2- \beta^2
 \een
from $C$ to $ \beta$. To this end, employing the known explicit
form of $\kappa$ we may re-write
 \be
C=C(\sigma,\tau,n)= (\sigma\alpha+2n+1) (\sigma\alpha+2n+1 +2\tau
\beta).
  \ee
This formula is linear in $\tau \beta$ and, hence, its easy
inversion defines the desirable state-dependent quantity $ \beta=
\beta(\sigma,\tau,n)$ as an elementary function of the constant
$C$.  The whole new energy spectrum acquires the closed form
 \be
E=E(\sigma,\tau,n)=A+B+\frac{1}{4}\,
\left [
\sigma\alpha+2n+1-\frac{A+B}{\sigma\alpha+2n+1}
\right ]^2.
 \ee
Our construction is complete.  The range of the quantum numbers
$n, \ \sigma$ and $\tau$ remains the same as above.

In the light of our new result we may now split the whole family
of the exactly solvable ${\cal PT}$ symmetric models in the two
distinct categories.  The first one ``lives" on the real line
and may be represented or illustrated not only by the popular
Laguerre-solvable harmonic oscillator  \cite{PTHO} but also by
our initial P\"{o}schl-Teller Jacobi-solvable force
(\ref{PTex}).

The second category requires a narrow arch-shaped path of
integration which all lies confined within a vertical strip.  It
contains at last both the Laguerre and Jacobi solutions.  The
former ones may be represented by the complex Morse model of ref.
\cite{Morse}.  Our present new Hulth\'en example offers its Jacobi
solvable counterpart.  The scheme becomes, in a way, complete.

The less formal difference between the two categories may be also
sought in their immediate physical relevance.  Applications of the
former class may be facilitated by a limiting transition which is
able to return them back on the usual real line.  In contrast, the
second category may rather find its most useful place in the
methodical considerations concerning, e.g., field theories and
parity breaking \cite{Bentri}.  Within the quantum mechanics
itself an alternative approach to the second category might also
parallel studies \cite{Cannata} of the ``smoothed" square wells in
the non-Hermitian setting.

In the conclusion let us recollect that the ${\cal PT}$ symmetry
of a Hamiltonian replaces and, in a way, {generalizes} its usual
hermiticity. This is the main reason why there existed a space for
a new solvable model among the singular interactions.  At the same
time, an exactly solvable example with another, intermediate
(i.e., hyperbola-shaped) arc of coordinates remains still to be
discovered.  Indeed, this type of a deformed contour has only been
encountered in the quasi-solvable (i.e., partially numerical)
model of ref. \cite{quartic}) and in the general unsolvable forces
studied by several authors by means of the perturbative
\cite{Caliceti}, numerical \cite{Guardiola} and WKB
\cite{Bender,WKB} approximation techniques.

\section*{Acknowledgement}

Partially supported by the grant Nr. A 1048004 of the Grant
Agency of the Academy of Sciences of the Czech Republic.

\end{document}